\documentclass[english,USenglish,oneside,twocolumn]{article}
\usepackage[T1]{fontenc}
\usepackage[utf8]{luainputenc}
\usepackage{float}
\usepackage{graphicx}
\usepackage{esint}

\makeatletter

\floatstyle{ruled}
\newfloat{algorithm}{tbp}{loa}
\providecommand{\algorithmname}{Algorithm}
\floatname{algorithm}{\protect\algorithmname}

\usepackage{fullpage}
\date{}
\@ifundefined{showcaptionsetup}{}{%
 \PassOptionsToPackage{caption=false}{subfig}}
\usepackage{subfig}
\makeatother

\usepackage{babel}
\begin{document}

\title{Untraceable VoIP Communication based on DC-nets}

\author{Christian Franck and Uli Sorger}
\maketitle
\begin{abstract}
Untraceable communication is about hiding the identity of the sender
or the recipient of a message. Currently most systems used in practice
(e.g., TOR) rely on the principle that a message is routed via several
relays to obfuscate its path through the network. However, as this
increases the end-to-end latency it is not ideal for applications
like Voice-over-IP (VoIP) communication, where participants will notice
annoying delays if the data does not arrive fast enough.

We propose an approach based on the paradigm of Dining Cryptographer
networks (DC-nets) that can be used to realize untraceable communication
within small groups. The main features of our approach are low latency
and resilience to packet-loss and fault packets sent by malicious
players. We consider the special case of VoIP communication and propose
techniques for a P2P implementation. We expose existing problems and
sketch possible future large-scale systems composed of multiple groups.

\end{abstract}

\section{Introduction}

A few decades ago telecommunications were mainly realized using circuit
switching, such that an electrical signal carrying the audio information
could be sent from one correspondent to another. Nowadays it is more
common that speech is digitized and then sent in small packets over
a private network or over the internet.

In such real-time VoIP communication the quality of the user experience
is dependent on the latency of the network. While some package-loss
can be tolerated, the one-way end-to-end latency should ideally not
exceed 150ms \cite{itu_g114}. To this effect a stream of small packets
is sent at a fast pace, for instance around 50 packets per second
\cite{BandwidthVOIP}.

We are interested to have a system in which two users can communicate
with each other via VoIP anonymously to the others. That means these
two users know that they are speaking to each other, but nobody else
can infer who is communicating with whom. 

One approach would be to use an anonymisation system like the well-known
TOR network \cite{dingledine2004tsg}, or alternatively one of the
more recently proposed systems like Drac \cite{danezis2010text} or
Herd \cite{le2015herd}. However, in all these systems the security
is based on the principle that the message is relayed several times
on its way through the network, so that it is hard to distinguish
who is the sender or the recipient of a message. One problem of this
is that users have to trust the operators of intermediary relays and
mixes that they are honest and do not disclose any information. Another
problem is that this relaying increases the latency and the jitter,
which reduces the quality of phone calls or video-conferences. 

Another approach would be to use Dining Cryptographer Networks (DC-nets).
Here, messages are not relayed but multiple players simultaneously
send ciphertexts to the recipient, and then the recipient combines
these ciphertexts to obtain a message. The recipient knows that one
of the players must have sent the message, but he is not be able to
distinguish which one. As opposed to relay based systems like TOR
and mixing, the latency is lower, no central authority is required
and might therefore be interesting for applications which require
a low latency and for P2P scenarios.

However, in classical DC-nets the recipient can only recover the message
once he has obtained all the ciphertexts. Furthermore, all these ciphertexts
must be correct; if a malicious player provides faulty data the recipient
cannot recover the message. Thus, if packets are late, lost or faulty,
then no useful information is transmitted. 

\makeatletter
\renewcommand{\algorithmname}{Protocol} 
\makeatother

In this paper, we propose a protocol that is based on the paradigm
of DC-nets but adapted for VoIP streaming. It enables two players
out of a group of $n$ players to communicate without disclosing to
anybody else that they are communicating. Our protocol is resilient
to packet-loss and to faulty packets, and can be implemented using
lightweight cryptography. We study the performance, discuss practical
implementations and propose possible future work. 

The paper is organized as follows. In Section~2 we describe our model
and our security requirements. In Section~3, we derive our new protocol.
In Section~4, we discuss practical considerations, e.g. an implementation
as a P2P protocol. In Section~5 we consider the performance. In Section~6
we sketch possible future work, and in Section~7 we review related
work. In Section~8 we conclude with some remarks.

\section{Model and Definitions \label{sec:Model-and-Definitions}}

\begin{figure}
\begin{centering}
\includegraphics[scale=0.5]{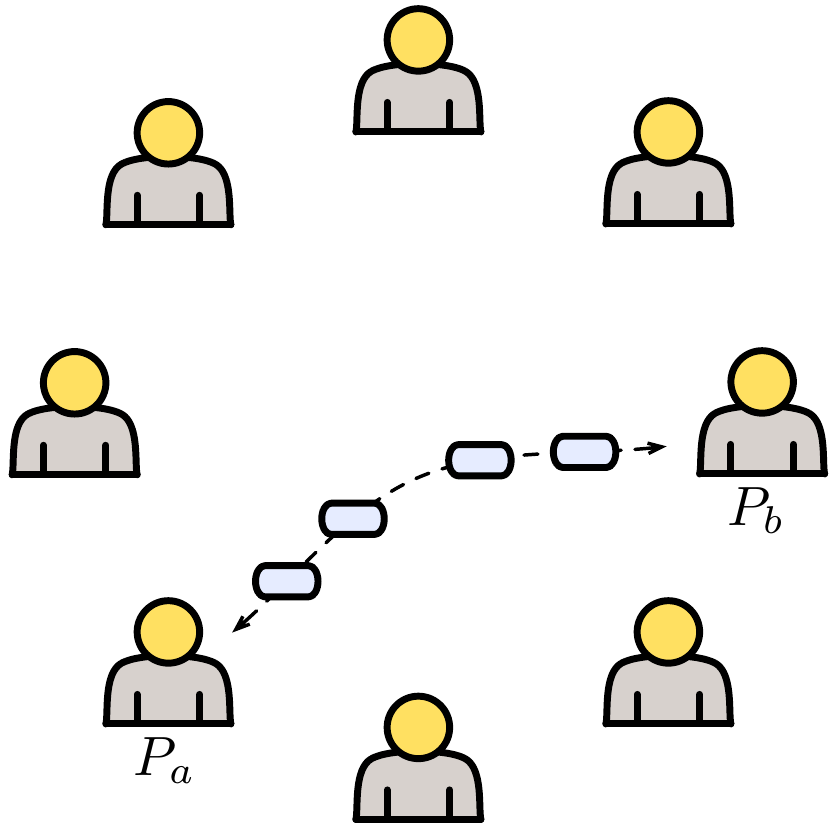}
\par\end{centering}

\caption{We want to enables two players $P_{a}$ and $P_{b}$ out of a group
of $n$ players $P_{1},...,P_{n}$ to communicate with VoIP in such
a way that nobody except them can distinguish which pair of the $n\cdot(n-1)$
possible pairs of players is communicating. }
\end{figure}

\begin{figure*}
\begin{centering}
\hfill{}\subfloat[Collection phase. Each player $P_{1},...,P_{n}$ sends a ciphertext
to the aggregator $A$. The ciphertexts of players $P_{a}$ and $P_{b}$
encode respectively the messages $m_{a}$ and $m_{b}$, whereas the
other ciphertexts are 'empty'. The aggregator $A$ aggregates the
received ciphertexts.]{\begin{centering}
\includegraphics[scale=0.5]{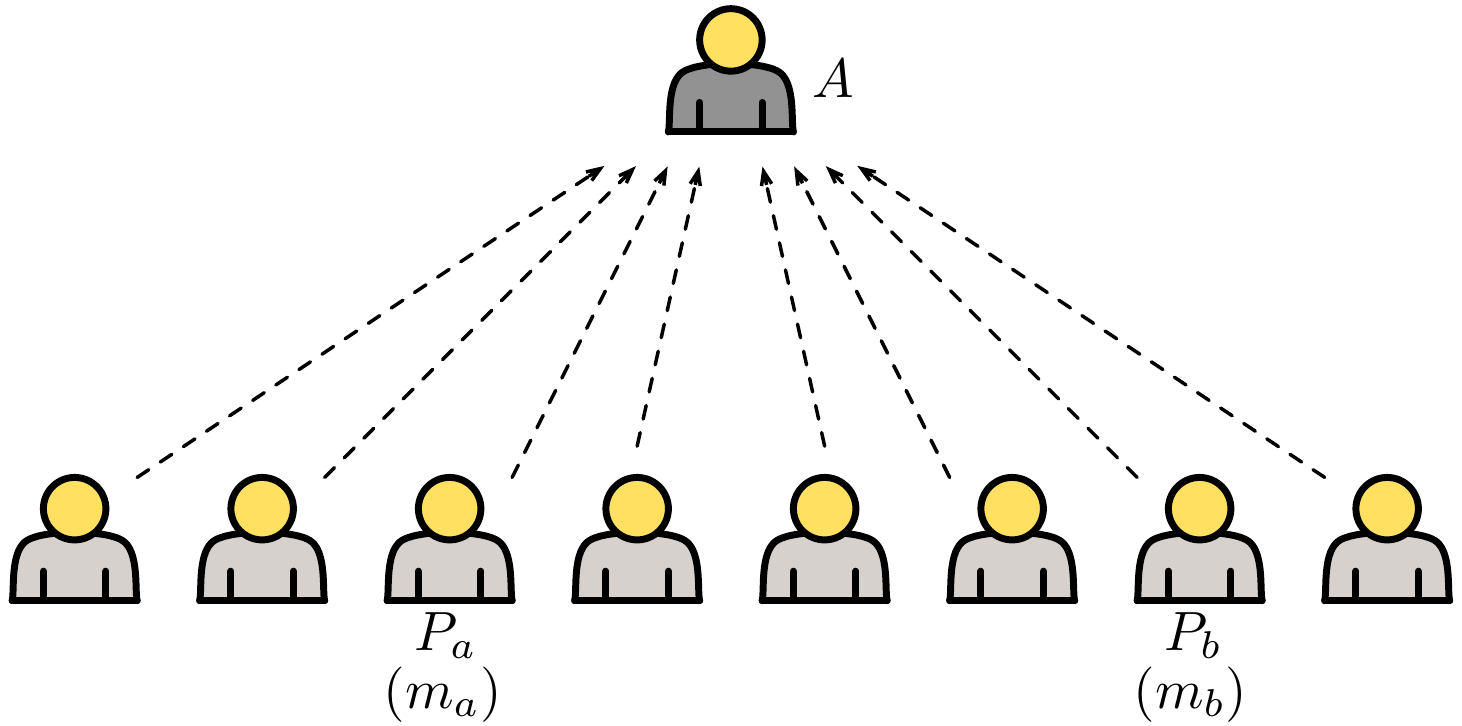}
\par\end{centering}

} \hfill{}\hfill{} \subfloat[Broadcast phase. The aggregator $A$ sends the result of the aggregation
back to all players $P_{1},...,P_{n}$. From this the players $P_{a}$
and $P_{b}$ can compute $m_{b}$ and $m_{a}$. ]{\begin{centering}
\includegraphics[scale=0.5]{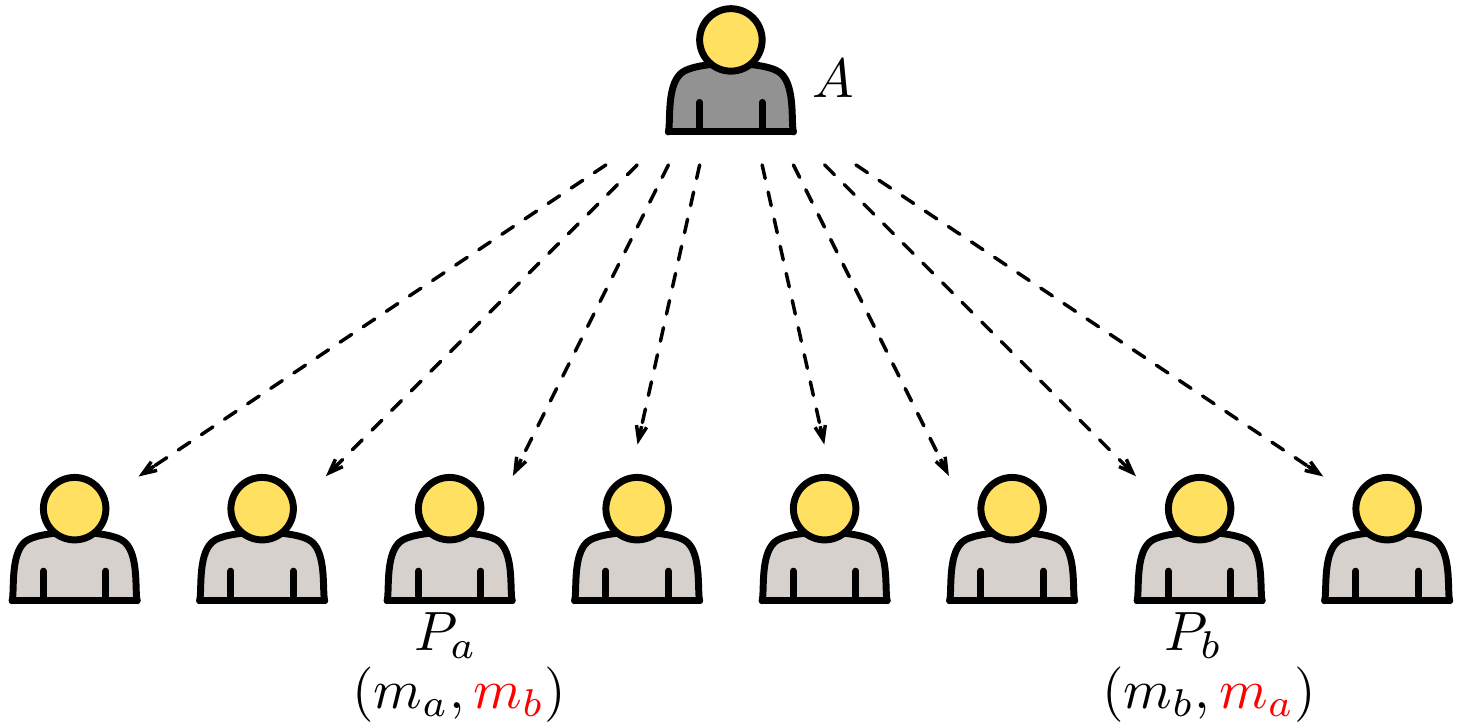}
\par\end{centering}

}\hfill{}
\par\end{centering}

\caption{We consider a system with $n$ players $P_{1},...,P_{n}$ and an aggregator
$A$, in which each transmission round comprises a sending and a broadcast
phase.}
\end{figure*}
We consider the case where two players want to communicate and stay
anonymous w.r.t. to the others, so our requirements are different
from those of the original DC-net protocol. In the original DC protocol
the recipient does not know the identity of the sender and the sender
does not know the identity of the recipient, but in our model we assume
that the two correspondent know each other's identity.

\subsection{Model}

We assume a communication network with $n$ players $P_{1},...,P_{n}$
and an aggregator $A$, and we consider the problem where two of these
players, $P_{a}$ and $P_{b}$, want to communicate with each other
such that nobody else can see that they are communicating.

After an appropriate setup, the messages are exchanged in successive
transmission rounds which consist of two phases:
\begin{enumerate}
\item A collection phase where each player send data to the aggregator.
\item A broadcast phase during which the aggregator sends the aggregated
data back to the participants.
\end{enumerate}
This star topology is a well-known scheme for implementing a DC-net
in networks without physical broadcast.

\subsection{Security Properties}

In this section we define the properties of correctness and privacy
we want to achieve.

\paragraph*{Correctness}

We say that a protocol is correct if, assuming the players $P_{a}$,
$P_{b}$ and the aggregator $A$ participate correctly, it holds that:
\begin{itemize}
\item if the message of $P_{a}$ encoding $m_{a}$ reaches the aggregator
$A$ in time and the result forwarded by the aggregator $A$ reaches
$P_{b}$, then $P_{b}$ can compute $m_{a}$; and 
\item if the message of $P_{b}$ encoding $m_{b}$ reaches the aggregator
$A$ in time and the result forwarded by the aggregator reaches $P_{a}$,
then $P_{a}$ can compute $m_{b}$. 
\end{itemize}
Thus, if the players $P_{a}$, $P_{b}$ and the aggregator $A$ behave
correctly and their packets arrive timely, then $P_{b}$ and $P_{a}$
can respectively recover $m_{a}$ and $m_{b}$.

\paragraph*{Privacy}

We say that a protocol is private if a computationally bounded adversary
controlling up to $n-3$ players (other than $P_{a}$ and $P_{b}$)
and the aggregator $A$ cannot determine who are the two correspondents
with a probability better than random guessing.

For a set of honest participants $H\subset\{P_{1},...,P_{n}\}$ there
are $c=\left|H\right|\cdot(\left|H\right|-1)\cdot2^{-1}$ different
pairs of correspondents. Thus a protocol is private if the adversary
has a probability of guessing correctly the pair of correspondents
which is not better than a random guess, i.e., with probability $1/c$.

\section{A Protocol for Untraceable Streaming}

To start we consider a simple protocol that meets the requirements
from the previous section in an ideal situation and we progressively
adapt it to end up with a protocol that can be used in a real-world
scenario. 

In the description we assume that the setups for the protocols have
already been performed. Typically one of the correspondents will anonymously
provide the other players with the required data using an anonymous
channel akin to \cite{FranckMscThesis,corrigan2013proactively}.

\subsection{A Simple Protocol }

\begin{algorithm}
\medskip{}

\begin{enumerate}
\item \noindent \textbf{Setup}: $P_{1},...,P_{n}$ respectively have secret
keys $k_{1},...,k_{n}$ such that $k_{1}+...+k_{n}=0$. $P_{a}$ additionally
has $m_{a}$ and $P_{b}$ additionally has $m_{b}$.
\item \noindent \textbf{Collection phase}: Every $P_{i}$ for $i\ne a,b$
sends $O_{i}=k_{i}$ to $A$. $P_{a}$ sends $O_{a}=k_{a}+m_{a}$
to $A$. $P_{b}$ sends $O_{b}=k_{b}+m_{b}$ to $A$. 
\item \noindent \textbf{Broadcast phase}: $A$ broadcasts $X=O_{1}+...+O_{n}$
to $P_{1},...,P_{n}$. $P_{a}$ computes $m_{b}=X-m_{a}$ and $P_{b}$
computes $m_{a}=X-m_{b}$.
\end{enumerate}
\noindent \raggedright{}\caption{A simple protocol based on DC-nets.}
\label{algo:1}
\end{algorithm}
We assume that during the execution of the protocol 
\begin{itemize}
\item the number of players remains constant,
\item no messages are lost (or arrive too late), and
\item all players are honest.
\end{itemize}

In this situation two players $P_{a}$ and $P_{b}$ can communicate
with each other in full-duplex using the classic DC-net principle.
This approach is described in detail in Protocol~\ref{algo:1}. All
players are initially provided with secret keys $k_{1},...,k_{n}$
that sum up to $0$, so that the keys will vanish when all ciphertexts
are aggregated. So if both $P_{a}$ and $P_{b}$ send during the same
round, the result forwarded by the aggregator is the sum of their
messages. By subtracting their own message from this result they can
recover each other's message. 

The previous system obviously fails, if only one user does not send
anything or if his packet is lost. This seems to be a very strong
restriction of the protocol.

\subsection{A Packet-Loss Resilient Protocol\label{sub:A-Packet-Loss-Resilient}}

The problem of packet-loss is that if the aggregator does not receive
all the packets, he can only make the sum $X$ over a strict subset
of $O_{1},...,O_{n}$. This means that the keys will not cancel and
$P_{a}$ and $P_{b}$ cannot recover the messages like in the previous
scenario. 

To overcome this problem we can modify the previous protocol as shown
in Protocol~\ref{protocol:2}. During the initialization $P_{a}$
and $P_{b}$ are provided with the keys $k_{1},...,k_{n}$ of all
players. Further the aggregator does not only broadcast the sum $X$,
but also a list $L$ informing which packets he received. Said list
$L$ informs the players $P_{a}$ and $P_{b}$ about which keys are
included in the partial sum and since they know all the keys, they
can subtract them from $X$ and recover the messages. 

\begin{algorithm}
\medskip{}

\begin{enumerate}
\item \textbf{Setup}: $P_{1},...,P_{n}$ respectively have secret keys $k_{1},...,k_{n}$.
$P_{a}$ additionally has $m_{a}$ and $P_{b}$ additionally has $m_{b}$,
and both know all $k_{1},...,k_{n}$.
\item \textbf{Collection phase}: Each $P_{i}$ for $i\ne a,b$ sends $O_{i}=k_{i}$
to $A$. $P_{a}$ sends $O_{a}=k_{a}+m_{a}$ to $A$. $P_{b}$ sends
$O_{b}=k_{b}+m_{b}$ to $A$. $A$ receives $O_{i}$ for $i\in L\subseteq\{1,...,n\}$.
\item \textbf{Broadcast phase}: $A$ broadcasts $(L,X)$ to $P_{1},...,P_{n}$,
where $X=\sum_{i\in L}O_{i}$. If $b\in L$ then $P_{a}$ computes
$m_{b}$; with $m_{b}=X-\sum_{i\in L}k_{i}-m_{a}$ in case $a\in L$
or otherwise with $m_{b}=X-\sum_{i\in L}k_{i}$. If $a\in L$ then
$P_{b}$ computes $m_{a}$; with $m_{a}=X-\sum_{i\in L}k_{i}-m_{b}$
in case $b\in L$ or otherwise with $m_{a}=X-\sum_{i\in L}k_{i}$. 
\end{enumerate}
\caption{A packet-loss resilient protocol.}
\label{protocol:2}
\end{algorithm}

This leaves us with the problem of users who deliberately send faulty
packets to disrupt the communication. Such a case should be caught
and the corresponding packets should be dropped. To identify such
packets, we propose the following protocol.

\subsection{A Protocol Resilient to Lost and Faulty Packets}

The problem is that if a malicious player $P_{i}$ sends a random
value instead of $k_{i}$ then $P_{a}$ and $P_{b}$ who expect $k_{i}$
will not be able to properly extract the messages $m_{a}$ and $m_{b}$
from $X$ anymore.

In order to protect against such malicious players it is obvious the
aggregator must be able to distinguish if a received packet is correct.
However the aggregator should not be able to distinguish which packets
contain messages. Thus every player must include a proof that the
submitted data is correct, and the aggregator must be able to verify
this proof without gaining any other information from it. This means
that a player $P_{i}\notin\{P_{a},P_{b}\}$ must be able to prove
that $O_{i}=k_{i}$, and $P_{a}$ and $P_{b}$ must keep the freedom
to send $O_{a}=k_{a}+m_{a}$ and $O_{b}=k_{b}+m_{b}$. 

An elegant way to achieve this is to bind each player $P_{i}$ to
his key $k_{i}$ using a trapdoor commitment, where the secret trapdoor
information $\alpha$ is only known to $P_{a}$ and $P_{b}$. Then
each player $P_{i}\notin\{P_{a},P_{b}\}$ can only open the commitment
to the value $k_{i}$, but $P_{a}$ and $P_{b}$ who know $\alpha$
can open their commitments to any value they like, that is to $k_{a}+m_{a}$
and $k_{b}+m_{b}$. 

In our description we use Pedersen commitments \cite{Pedersen91}
which are of the form $c=g^{r}h^{k}$, and we assume that the secret
$\alpha=\log_{g}h$ is only known to $P_{a}$ and $P_{b}$.

As shown in Protocol~\ref{protocol:3}, during the setup the aggregator
is provided with a commitment for each expected $O_{i}$, and each
player $P_{i}$ is provided with the corresponding secret pairs $(k_{i},r_{i})$.
Then, during the collection phase, each participant $P_{i}\notin\{P_{a},P_{b}\}$
must send $(k_{i},r_{i})$ to the aggregator, since without $\alpha$
he cannot find any other pair $(k'_{i},r'_{i})$ that corresponds
to the commitment. $P_{a}$ and $P_{b}$ can use $\alpha$ to compute
valid pairs $(k_{a}+m_{a},r'_{a})$ and $(k_{b}+m_{b},r'_{b})$. The
aggregator verifies for each received pair if it corresponds to the
commitment and rejects pairs that do not. Thus only valid $k_{i}$s
are used to compute $X$.

\begin{algorithm}
\medskip{}

\begin{enumerate}
\item \textbf{Setup}: $P_{1},...,P_{n}$ respectively have secret value
pairs $(k_{1},r_{1}),...,(k_{n},r_{n})$. $P_{a}$ additionally has
$m_{a}$ and $P_{b}$ additionally has $m_{b}$, and both know all
$(k_{1},r_{1}),...,(k_{n},r_{n})$. $A$ is provided with $c_{1},...,c_{n}$
where $c_{i}=g^{r_{i}}h^{k_{i}}$. Only $P_{a}$ and $P_{b}$ know
$\log_{g}h$.
\item \textbf{Collection phase}: Each $P_{i}$ sends $(O_{i},s_{i})$ to
$A$. Each $P_{i}$ for $i\ne a,b$ uses $O_{i}=k_{i}$ and $s_{i}=r_{i}$.
$P_{a}$ uses $O_{a}=k_{a}+m_{a}$ and $s_{i}=r_{i}-m_{a}\cdot\log_{g}h$.
$P_{b}$ uses $O_{b}=k_{b}+m_{b}$ and $s_{i}=r_{i}-m_{b}\cdot\log_{g}h$.
$A$ receives $(O_{i},s_{i})$ where additionally $g^{s_{i}}h^{O_{i}}=c_{i}$
holds for $i\in L\subseteq\{1,...,n\}$.
\item \textbf{Broadcast phase}: $A$ broadcasts $(L,X)$ to $P_{1},...,P_{n}$,
where $X=\sum_{i\in L}O_{i}$. If $b\in L$ then $P_{a}$ computes
$m_{b}$; with $m_{b}=X-\sum_{i\in L}k_{i}-m_{a}$ in case $a\in L$
or otherwise with $m_{b}=X-\sum_{i\in L}k_{i}$. If $a\in L$ then
$P_{b}$ computes $m_{a}$; with $m_{a}=X-\sum_{i\in L}k_{i}-m_{b}$
in case $b\in L$ or otherwise with $m_{a}=X-\sum_{i\in L}k_{i}$. 
\end{enumerate}
\caption{A packet-loss resilient protocol with verification.}
\label{protocol:3}
\end{algorithm}

This scenario is good if there is only one transmission round, but
the anonymous sending of $c_{1},...,c_{n}$ to $A$ is expensive and
does not scale well to multiple rounds. Therefore we need a more efficient
way to provide the aggregator $A$ with means for verifying the data
from the participants when there are multiple transmission rounds.

\subsection{A Protocol Resilient to Packet-Loss and Malicious Players for Multiple
Rounds\label{sub:A-Protocol-Resilient}}

\begin{figure*}
\begin{centering}
\includegraphics[scale=0.5]{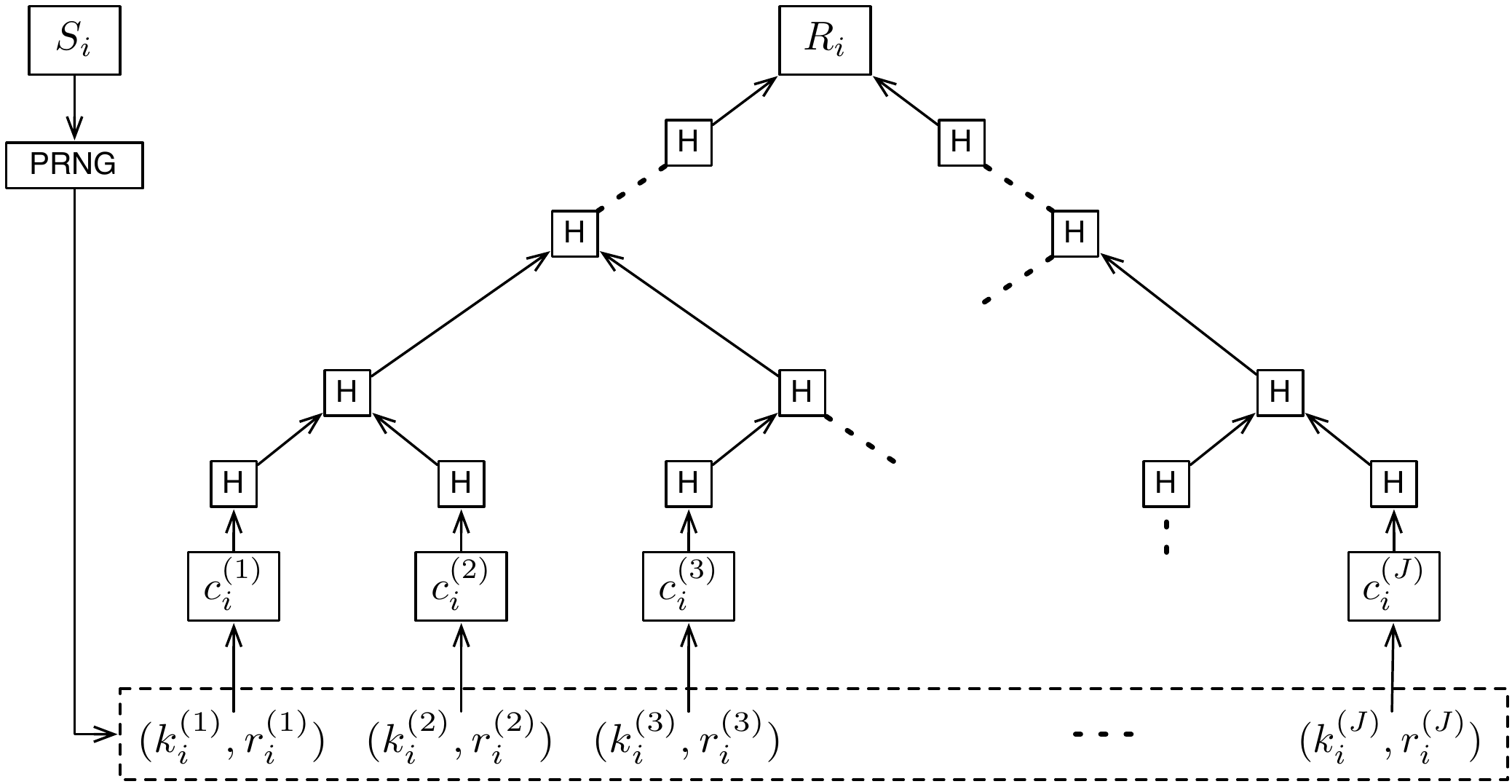}
\par\end{centering}

\caption{A Merkle tree $T_{i}$ for the sequence of commitments $c_{i}^{(1)},...,c_{i}^{(J)}$
of player $P_{i}$. The sequence of the secret keys $(k_{i}^{(1)},r_{i}^{(1)}),...,(k_{i}^{(J)},r_{i}^{(J)})$
can be pseudo-randomly generated from a secret seed $S$. The player
$P_{i}$ can prove that the commitment $c_{i}^{(j)}$ is at the position
$j$ in the sequence.}
\label{Fig:Merkle}
\end{figure*}

In order to extend the previous protocol to multiple transmission
rounds, we propose to use Merkle trees~\cite{merkle1989certified}.
For each player $P_{i}$ we use a Merkle tree $T_{i}$ that allows
to verify that a given commitment is valid for a given round. It is
then not necessary anymore to provide the aggregator with a commitment
for each round, but it is sufficient to provide the aggregator with
the roots of the Merkle trees. As illustrated in Figure~\ref{Fig:Merkle},
such a Merkle tree $T_{i}$ can be constructed from a sequence $(k_{i}^{(1)},r_{i}^{(1)}),...,(k_{i}^{(J)},r_{i}^{(J)})$,
which can be derived from a secret seed $S_{i}$. 

So what changes compared to the preceding protocol is that the aggregator
is provided with the roots of the Merkel trees instead of the commitments,
and each player $P_{i}$ is provided with a secret seed $S_{i}$ that
corresponds to a pseudorandom sequence. Then, during the transmission
phase each player $P_{i}$ does not only send $(k_{i}^{(j)},r_{i}^{(j)})$
but $(k_{i}^{(j)},r_{i}^{(j)},Z_{i}^{(j)})$ where $Z_{i}^{(j)}$
is a proof the commitment corresponding to $(k_{i}^{(j)},r_{i}^{(j)})$
is the right one for round $j$ (i.e., that it is at position $j$
in the sequence). The aggregator computes a commitment and verifies
using $Z_{i}^{(j)}$ that it is correct. A detailed description of
the protocol is shown in Protocol~\ref{protocol:4}.

It is easy to see that this protocol satisfies the properties of correctness
and privacy defined in Section~\ref{sec:Model-and-Definitions}.

\begin{algorithm}
\medskip{}

\begin{enumerate}
\item \noindent \textbf{Setup}: For each round $j\in\{1,...,J\}$, $P_{1},...,P_{n}$
respectively have secret value pairs $(k_{1}^{(j)},r_{1}^{(j)}),...,(k_{n}^{(j)},r_{n}^{(j)})$.
$P_{a}$ additionally has $m_{a}^{(j)}$ and $P_{b}$ additionally
has $m_{b}^{(j)}$, and both know all $(k_{1}^{(j)},r_{1}^{(j)}),...,(k_{n}^{(j)},r_{n}^{(j)})$.
$A$ is provided with $R_{1},...,R_{n}$ the roots of a merkle trees
$T_{1},...,T_{2}$ constructed from $(c_{1}^{(1)},...,c_{1}^{(J)}),...,(c_{n}^{(1)},...,c_{n}^{(J)})$
where $c_{i}^{(j)}=g^{r_{i}^{(j)}}h^{k_{i}^{(j)}}$. Only $P_{a}$
and $P_{b}$ know $\log_{g}h$.
\item \noindent \textbf{Collection phase} (round $j$): Each $P_{i}$ sends
$(O_{i}^{(j)},s_{i}^{(j)},z_{i}^{(j)})$ to $A$. Each $P_{i}$ for
$i\ne a,b$ uses $O_{i}^{(j)}=k_{i}^{(j)}$ and $s_{i}^{(j)}=r_{i}^{(j)}$.
$P_{a}$ uses $O_{a}^{(j)}=k_{a}^{(j)}+m_{a}^{(j)}$ and $s_{i}^{(j)}=r_{i}^{(j)}-m_{a}^{(j)}\cdot\log_{g}h$.
$P_{b}$ uses $O_{b}^{(j)}=k_{b}^{(j)}+m_{b}^{(j)}$ and $s_{i}^{(j)}=r_{i}^{(j)}-m_{b}^{(j)}\cdot\log_{g}h$.
Further, $z_{i}^{(j)}$ is a proof that $c_{i}^{(j)}=g^{r_{i}^{(j)}}h^{k_{i}^{(j)}}$
is in the Merkle tree $T_{i}$ at position $j$. $A$ receives $(O_{i}^{(j)},s_{i}^{(j)},z_{i}^{(j)})$
where $z_{i}^{(j)}$ proves that $c_{i}^{(j)}=g^{s_{i}^{(j)}}h^{O_{i}^{(j)}}$
is at position $j$ in $T_{i}$, for $i\in L^{(j)}\subseteq\{1,...,n\}$.
\item \noindent \textbf{Broadcast phase} (round $j$): $A$ broadcasts $(L^{(j)},X^{(j)})$
to $P_{1},...,P_{n}$, where $X^{(j)}=\sum_{i\in L^{(j)}}O_{i}^{(j)}$.
If $b\in L^{(j)}$ then $P_{a}$ computes $m_{b}^{(j)}$; with $m_{b}^{(j)}=X^{(j)}-\sum_{i\in L^{(j)}}k_{i}^{(j)}-m_{a}^{(j)}$
in case $a\in L^{(j)}$ or otherwise with $m_{b}^{(j)}=X^{(j)}-\sum_{i\in L^{(j)}}k_{i}^{(j)}$.
If $a\in L^{(j)}$ then $P_{b}$ computes $m_{a}^{(j)}$; with $m_{a}^{(j)}=X^{(j)}-\sum_{i\in L^{(j)}}k_{i}^{(j)}-m_{b}^{(j)}$
in case $b\in L^{(j)}$ or otherwise with $m_{a}^{(j)}=X^{(j)}-\sum_{i\in L^{(j)}}k_{i}^{(j)}$. 
\end{enumerate}
\caption{A packet-loss resilient protocol with verification and multiple rounds.}
\label{protocol:4}
\end{algorithm}

\subsection{Variants of the Protocol}

In order to recover the messages in presence of packet-loss we proposed
in Section~\ref{sub:A-Packet-Loss-Resilient} that the aggregator
should send the list of packets that have been received along with
the sum. This allows the receiver to directly compute the message.
If one can assume that only a few (e.g., 1 or 2) packets are lost
per round, one can also omit to send this list. The recipient can
then still recover the message by trying all the possible combinations
of missing packets. This way the packet length is reduced, however
at the cost of a more expensive computation at the recipient.

Similarly one could also completely omit the cryptographic proof and
go for a completely different mechanism. The players $P_{a}$ and
$P_{b}$ could, upon detection of problems, use the anonymous channel
from the setup and ask the aggregator to publish all the packets he
received during a problematic round. Since $P_{a}$ and $P_{b}$ know
all the keys, they would directly distinguish who sent a faulty packet,
and they could anonymously ban those players from the group. This
optimistic approach would lead to shorter packets, but as the latency
of the anonymous channel is expected to be high, the stream would
be interrupted for a non negligible amount of time.

\section{Practical considerations}

In this section we discuss some aspects to consider in a real implementation
of the protocol.

\paragraph{Channel Setup}

Concerning the setup of the channel, we assumed so far that the initialization
is performed anonymously by one of the correspondents $P_{a}$ or
$P_{b}$. This correspondent will provide all other players and the
aggregator with the the required data via an anonymous communication
channel. 

One way to implement such an anonymous channel is to use a DC-net.
However, such a DC-net must then be run periodically, since in general
it is not known in advance when a correspondent will want to talk
with another. The higher the frequency with which such a DC-net is
run, the better the user experience. But as each run consumes bandwidth,
one does not want to do this more often than necessary either. So
there is a tradeoff to be made between bandwidth and user experience
with this approach.

Another way to implement an anonymous channel is to use a relay based
approach like onion routing (e.g. TOR). Here the problem is that the
overall security provided by the system is only as strong as the weakest
link in the chain. The use of such a relay based approach would weaken
the overall security of the system.

\paragraph{Channel Termination}

In the protocol of Section~\ref{sub:A-Protocol-Resilient} the number
of rounds (and thus the length of the call) is fixed during the setup
of the channel. If a call ends earlier the correspondents can actively
terminate the call by notifying the other players via the same anonymous
channel that they used to do the initialization.

\paragraph{Load Distribution with P2P}

Especially for the aggregator the computational costs and the bandwidth
requirements and can rise to non-negligible levels, since they are
proportional to the number of players. For instance if all packets
are around 100 bytes and if 100 players send 50 packet per second,
the aggregator must aggregate $5000$ packets per second and has a
corresponding incoming and outgoing traffic of $4\,\mbox{Mbps}$.
Each participant would have an incoming an outgoing traffic of $40\,\mbox{kbps}$.

\begin{figure}
\subfloat[Implementation with a dedicated aggregator: In each round the same
aggregator processes the data from all the participants.]{\noindent \begin{centering}
\includegraphics[scale=0.48]{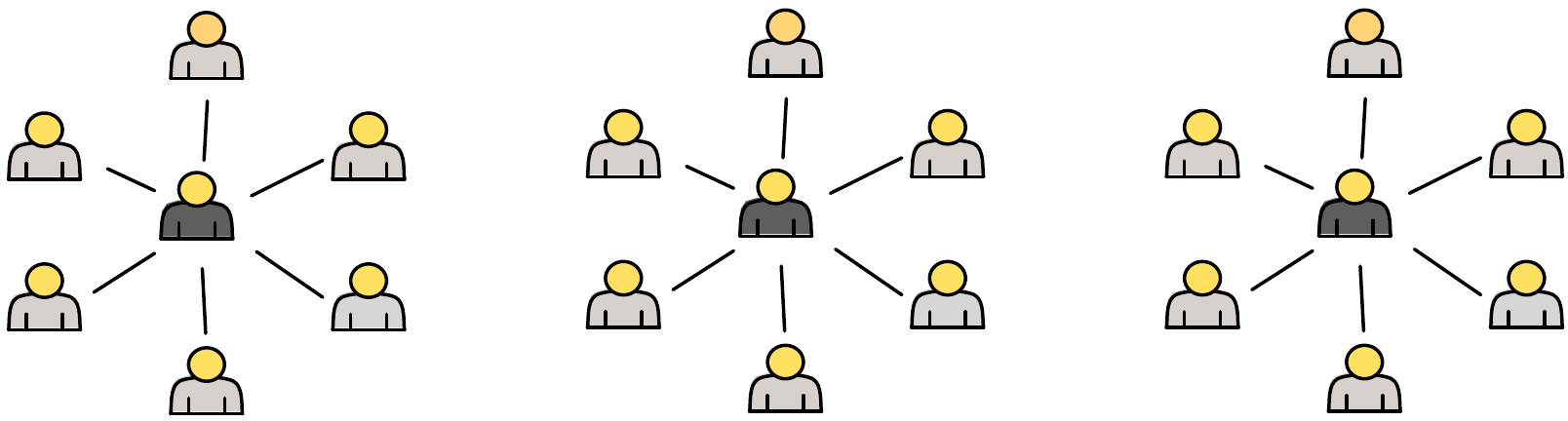}
\par\end{centering}

\label{Fig:p2p_a}}

\subfloat[Implementation as a distributed P2P system: In each round another
party plays the role of the aggregator. Advantageously this also mitigates
the damage that dishonest aggregator can cause.]{\noindent \begin{centering}
\includegraphics[scale=0.48]{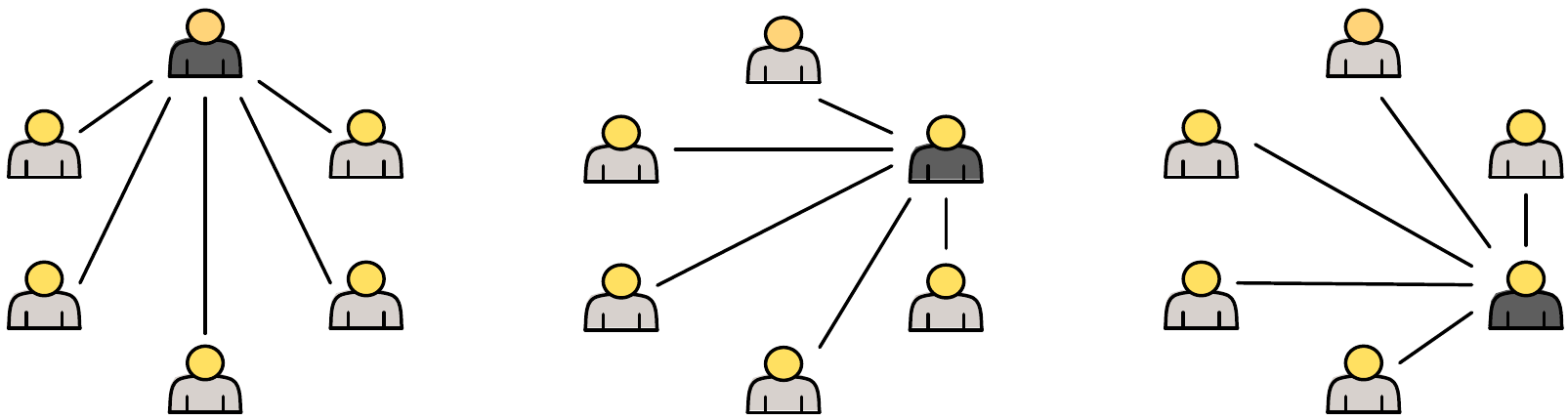}\bigskip{}

\par\end{centering}

\label{Fig:p2p_b}}

\caption{Centralized and distributed systems.}
\label{Fig:p2p_rotation}
\end{figure}

In a P2P system one is not obliged to have only one aggregator as
illustrated in Figure~\ref{Fig:p2p_a}, but the players can distribute
this load between all of them by successively have each one of them
play the role of the aggregator in a round robin fashion as illustrated
in Figure~\ref{Fig:p2p_b}. This way the load is more evenly distributed
and for the same setup as in the preceding example each player would
have of around $80\,\mbox{kbps}$ of ignoring and outgoing traffic.

\paragraph{Synchronization}

All players should send their packets such that they arrive at the
aggregator practically at then same time, in order to minimize the
overall latency of a transmission. The aggregator will only wait for
a certain period of time before aggregating the received data and
sending the result to the players. It is therefore important that
the clocks of the players are properly synchronized.

\paragraph{Cryptography}

The cryptographic assumption for the commitments and the hashtables
only needs to hold for a short time. It is therefore possible to use
a lower security parameter than for digital signatures that have to
be secure for decades.

\section{Performance}

In this section we consider the latency, the bandwidth and the computing
complexity for some setups.

\paragraph{Latency}

\begin{figure*}
\begin{centering}
\subfloat[Latency of a packet.]{\includegraphics[scale=0.37]{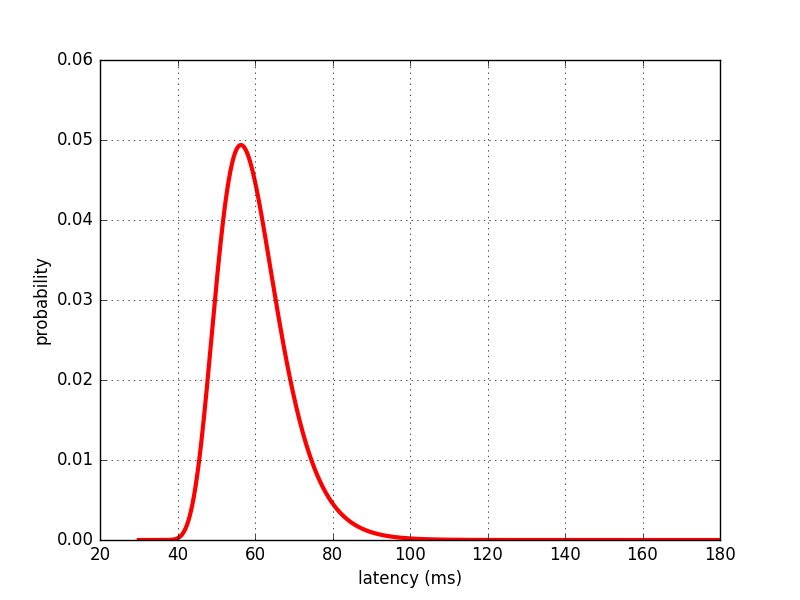}

\label{fig:latency_log_normal}}\subfloat[Average latency of the last packet.]{\includegraphics[scale=0.37]{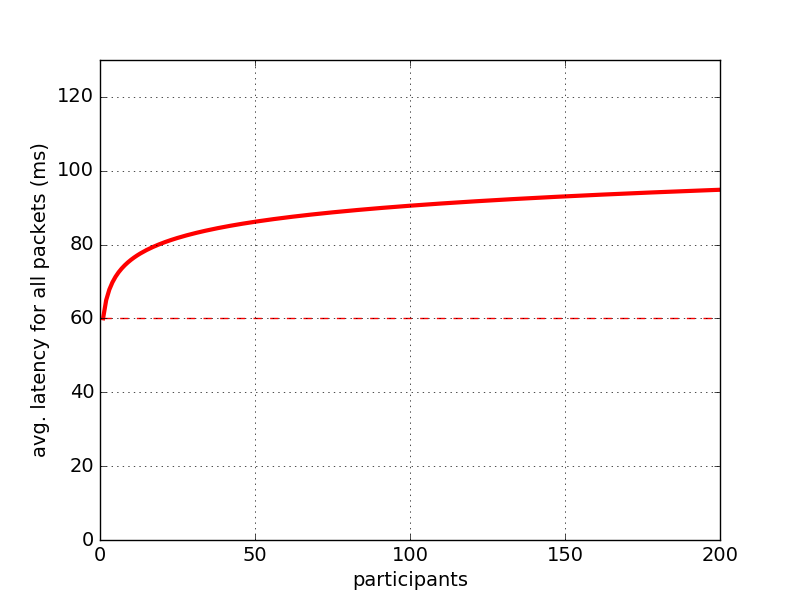}

\label{fig:cumul_latency}}
\par\end{centering}

\caption{Latency of arrival of the packets at the aggregator.}
\end{figure*}

Latency of packets in computer networks is often assumed to follow
a log-normal distribution, e.g. in \cite{huberman1997social,Sorger2011}.
This distribution defined by 
\[
Pr(t=x)=\frac{1}{\sqrt{2\pi s^{2}}}\cdot\exp\left(\frac{(\ln x-u)^{2}}{2s^{2}}\right)
\]
has a characteristic heavy tail, as shown shown in Figure~\ref{fig:latency_log_normal}.
If we assume an average $u=0.97$ and a standard error $s=0.06$,
then the average time the aggregator has to wait until all $n$ independent
ciphertexts have arrived is given by
\[
Pr(t<x)^{n}=\left(\int_{l=-\infty}^{x}Pr(t=l)\right)^{n}.
\]
where $n$ is the number of players. As shown in Figure~\ref{fig:cumul_latency}
the cumulated latency increases with the number of players. In our
case, we see that for $n=100$ players we already have a latency increase
of more than $30$ms.

\paragraph{Packet Loss}

\begin{figure*}
\begin{centering}
\includegraphics[scale=0.37]{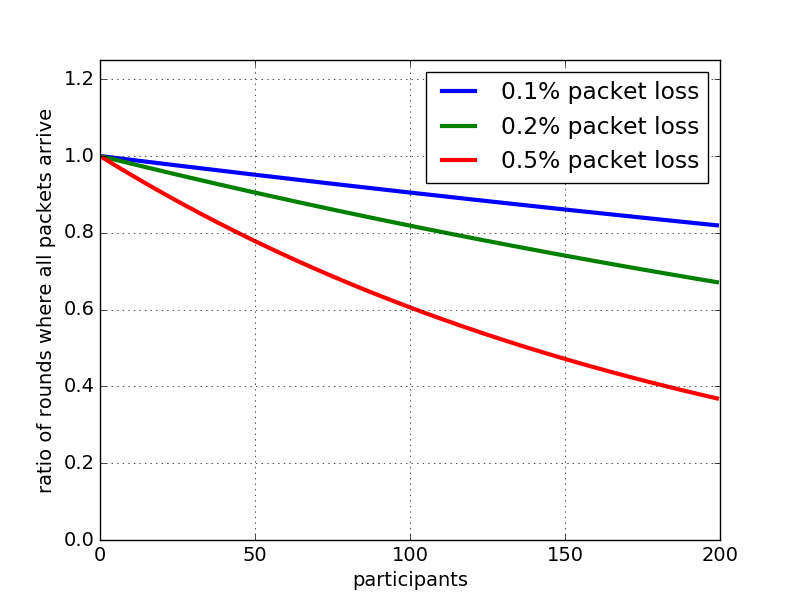}
\par\end{centering}

\caption{Ratio of rounds in which no packet is lost on its way to the aggregator.}
\label{packetloss}
\end{figure*}

Packet loss typically occurs in bursts and can be modeled using the
well known Gilbert–Elliott (GE) channel \cite{gilbert1960capacity,elliott1963estimates}.
We estimate the number of rounds during which no packet is lost on
its way to the aggregator, based on the probability $p$ that a packet
is lost. The probability that a packet is not lost is then $1-p$,
and the probability that no packet is lost is then 
\[
q=(1-p)^{n}.
\]
Figure~\ref{packetloss} illustrates the number of rounds during
which at least one packet does not make it to the aggregator for various
values of $p$.

\paragraph{Bandwidth}

During one transmission round each of the $n$ players sends a packet
to the aggregator, and the aggregator sends a packet back to each
player. The total number of packets per second $b(n)$ is thus proportional
to the number of players $n$. That is 
\[
b(n)=2\cdot\frac{n-1}{f},
\]
where $f$ is the number of rounds per second.

The load of the aggregator increases with the number of players. 

\begin{figure*}
\begin{centering}
\subfloat[Packets sent per participant.]{\includegraphics[scale=0.37]{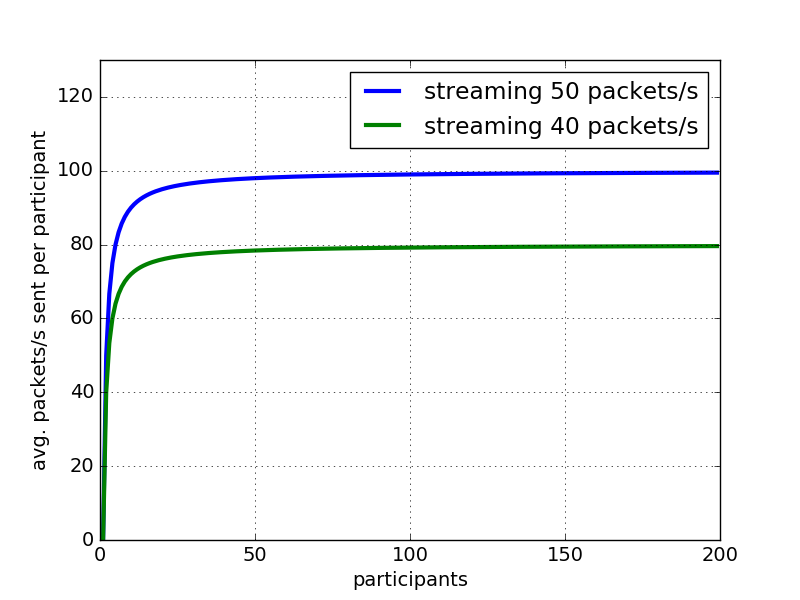}

\label{Fig_bandwidth_per_participant}}\subfloat[Total number of packets sent.]{\includegraphics[scale=0.37]{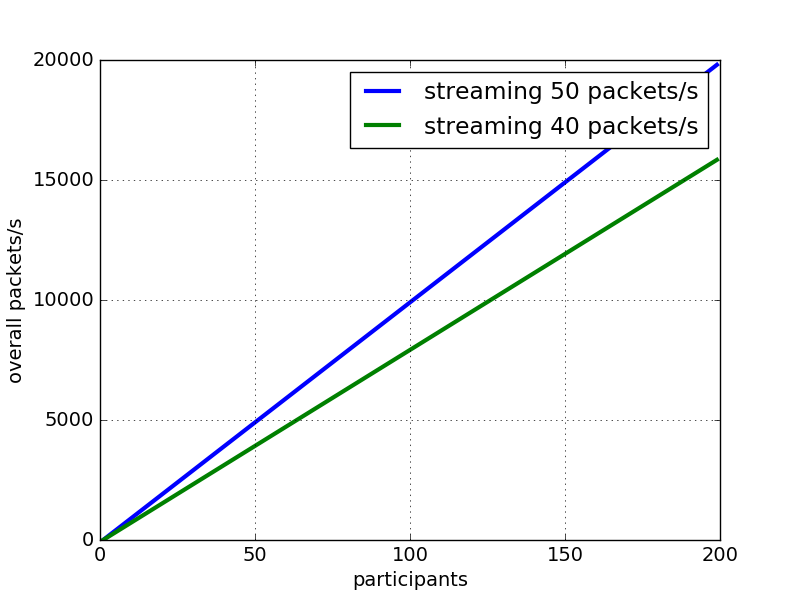}

\label{Fig_bandwidth_total}}
\par\end{centering}

\caption{Bandwidth}
\label{Fig:bandwidth}
\end{figure*}

In a P2P scenario where all players successively play the role of
the aggregator, the bandwidth usage is distributed evenly amongst
all players. Each player will perform like a normal player for n-1
rounds, and in one out of $n$ rounds he will not have to send anything,
but he will have to broadcast the aggregated data to the $n-1$ other
players. The bandwidth per player $p(n)$ is shown in Figure~\ref{Fig_bandwidth_per_participant}.
It can be computed with 
\[
p(n)=\frac{b}{n}=\frac{n-1}{n}\cdot\frac{2}{f}\sim\frac{2}{f}.
\]

\paragraph{Packet size}

Packets in our protocol are composed of two parts, the audio payload
on one hand and the cryptographic overhead on the other hand. 

The amount of audio data depends on the frequency of the packets,
on the quality (sampling frequency, compression rate) of the sound
and the number of sound channels (e.g., mono or stereo). For voice
transmission in mono this could be 50 packets with 60 bytes per second
per packet, but for high-end music in stereo it will be significantly
more. 

The amount of cryptographic data depends on the strength of the cipher
that is used. Since the cryptographic assumption only has to hold
during the communication one can use lightweight cryptographic primitives.


\section{Future Work}

There are basically two directions for future work, the improvement
of the protocol itself and the building of larger systems composed
of multiple groups.

\begin{figure}
\begin{centering}
\includegraphics[scale=0.48]{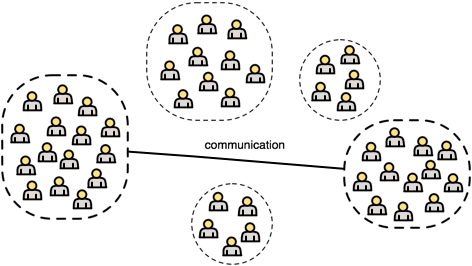}
\par\end{centering}

\caption{In an untraceable communication system with groups, there is a whole
group of potential correspondents on either side of the line. Nobody
except the real correspondents -- neither an external observer, nor
any other group member -- can distinguish who is communicating with
whom. }

\label{Figure:GlobalVision-1}
\end{figure}

\paragraph{Detection of Malicious Aggregators}

In this paper we assume that the aggregator is honest but curious.
This means that he will not drop packets, nor omit to send packets.
While he cannot send wrong results and remain undetected, he can just
disrupt a transmission round by just dropping data. A more powerful
malicious aggregator could however just ignore some of the packets
he receives, or he could deliberately not broadcast the aggregated
result to anybody. In a P2P setting the effect of such an aggregator
can be mitigated using the rotation principle proposed in section~X,
but ideally one would like to detect and to ban such aggregators from
the group.

\paragraph{Larger Systems with Multiple Groups }

Protocols based on DC-net do not scale to a very large number of participants,
as the bandwidth and the computational power used by the aggregator
are proportional to the number of participants. So the idea which
was already proposed in \cite{goel2003hsa} is to realize systems
composed of many small groups, as illustrated in Figure~\ref{Figure:GlobalVision-1}.
Only the correspondents will know that they are communicating, all
other players or observers cannot distinguish who is communicating. 

For example one could have on one side a group of 500 politicians
and on the other side a group of 500 journalists. When a politician
then talks to a journalist, it would only be possible to see that
one of the 500 politicians is talking to one of the 500 journalists,
but it would not be possible for anybody to distinguish which politician
is talking to which journalist. As there would be $500\cdot500=250000$
different possibilities, such a system would offer a fairly good protection.

There are many open questions, such as: How can we locate a given
participant within the system, if we do not know in which group he
is? How can we handle participants joining and leaving the system?
How can we ban malicious participants from the entire system?

\section{Related Work}

The Dining Cryptographers protocol was proposed in \cite{chaum1988dcp}
and further studied in \cite{bos1989ddd,BWPW91,waidner1990usa,waidner1989dcd}.
A first system composed of multiple DC-nets was proposed in \cite{goel2003hsa}. 

Computationally secure DC-net protocols with zero-knowledge verification
of the data have been proposed in \cite{golle2004dcr} and further
studied in \cite{FranckMscThesis,wolinsky2012dissent,corrigan2013proactively,Franck_DC_0924,corrigan2015riposte}. 

Recent group based communication systems include \cite{corrigan2015riposte,kwon2015riffle,le2015herd,danezis2010text}.
There have also been TOR \cite{dingledine2004tsg} extensions for
VoIP \cite{gegelonionphone} and for group communication \cite{yang2015mtor}.

\section{Concluding Remarks}

Starting from the classic DC-net paradigm we derived a protocol for
untraceable VoIP telephony that is resilient against packet-loss and
faulty packets. It enables two players within a larger group to communicate
with VoIP without anybody else being able to distinguish that they
are communicating. Further we discussed practical issues and showed
how to distribute the load in a P2P network. 

We consider this work a first step towards larger systems composed
of multiple groups so that can scale to a larger number of participants. 

\bibliographystyle{plain}
\bibliography{references}

\end{document}